\documentclass{emulateapj}

\usepackage{longtable}
\usepackage{amsmath}
\renewcommand{\thefootnote}{\fnsymbol{footnote}}

\begin{document}

\title{Direct Determination of the HF/H$_2$ Abundance Ratio in Interstellar Gas\footnote{Based on observations collected at the European Organisation for Astronomical Research in the Southern Hemisphere, Chile as part of program 089.C-0321}}

\author{Nick Indriolo\altaffilmark{1},
D.~A.~Neufeld\altaffilmark{1},
A.~Seifahrt\altaffilmark{2},
M.~J.~Richter\altaffilmark{3}
}

\altaffiltext{1}{Department of Physics and Astronomy, Johns Hopkins University, Baltimore, MD 21218}
\altaffiltext{2}{Department of Astronomy and Astrophysics, University of Chicago, Chicago, IL 60637}
\altaffiltext{3}{Department of Physics, University of California Davis, Davis, CA 95616}

\begin{abstract}

We report the first detection of the $v=1$--0, $R(0)$ ro-vibrational transition of HF at 2.499385~$\mu$m arising from interstellar gas.
The line is seen in absorption toward 3 background sources---HD~154368, Elias~29, and AFGL~2136~IRS~1---all of which have reported H$_2$ column densities determined from observations of H$_2$.  This allows for the first direct determination of the HF/H$_2$ abundance ratio.  We find values of $N({\rm HF})/N({\rm H}_2)=1.15\times10^{-8}$ and $0.69\times10^{-8}$ for HD~154368 and Elias~29, respectively.  The sight line toward AFGL~2136~IRS~1 also shows absorption from the $v=1$--0, $R(1)$ transition of HF, indicating warm, dense ($n_{\rm H}\gtrsim 10^9$~cm$^{-3}$) gas, likely very close to the central protostar.  Ascribing portions of the HF absorption to warm and cold gas, we find $N({\rm HF})/N({\rm H}_2)=(1.7$--$2.9)\times10^{-8}$ and $(0.33$--$0.58)\times10^{-8}$ for the two components, respectively.  Except for the warm component toward AFGL~2136~IRS~1, all observed HF/H$_2$ ratios are well below $N({\rm HF})/N({\rm H}_2)=3.6\times10^{-8}$, the value predicted if nearly all gas-phase fluorine is in the form of HF.  Models of fluorine chemistry that account for depletion onto grains are able to reproduce the results toward HD~154368, but not in the cold, dense gas toward AFGL~2136~IRS~1 and Elias~29.  Most likely, some combination of simplifying assumptions made in the chemical models are responsible for these discrepancies.
\end{abstract}


\section{INTRODUCTION} \label{section_intro}

\setcounter{footnote}{3}
\renewcommand{\thefootnote}{\arabic{footnote}}

Fluorine is unique among all elements in that its neutral, atomic form reacts exothermically with H$_2$ because the dissociation energy of HF is larger than that of H$_2$.  The ionization potential of F is larger than that of H, such that atomic fluorine is expected to remain in neutral form throughout most of the interstellar medium (ISM).  In regions with H$_2$, fluorine will convert to molecular form via the reaction
%In regions with substantial amounts of molecular hydrogen, chemical models predict that the majority of gas phase fluorine will convert to molecular form via the reaction
\begin{equation}
{\rm F} + {\rm H}_2 \rightarrow {\rm HF} + {\rm H}.
\label{rx_F_H2}
\end{equation}
HF is destroyed by photodissociation and ion-neutral reactions (with, e.g., C$^+$, Si$^+$, He$^+$, H$_3^+$).  While these processes---most importantly reactions with C$^+$ and photodissociation---can compete with reaction (\ref{rx_F_H2}) in the diffuse, primarily atomic ISM to keep a significant fraction of fluorine in atomic form, chemical models predict that in molecular gas (H$_2$/H $\gtrsim1$) HF becomes the dominant reservoir of fluorine \citep{neufeld2005,zhu2002}.
%Because destruction of HF via photodissociation and ion-neutral reactions (with, e.g., C$^+$, Si$^+$, He$^+$, H$_3^+$) is slow relative to reaction (\ref{rx_F_H2}), HF is predicted to be the dominant reservoir of fluorine in molecular gas \citep{neufeld2005,zhu2002}.

Interstellar HF was first discovered using the {\it Infrared Space Observatory} \citep[ISO;][]{kessler1996} targeting the $J=2$--1 transition at 121.6973~$\mu$m toward Sgr B2 \citep{neufeld1997}.  Given the large dipole moment and rotational constant of HF though, only in exceptional cases of radiative pumping (as is the case with Sgr B2) or at high densities ($n_{\rm H}\gtrsim 10^9$~cm$^{-3}$) is the $J=1$ level expected to be significantly populated.  In most regions of the ISM, nearly the entire population of HF is expected to be in the $J=0$ ground state, thus limiting the utility of the $J=2$--1 line.  It was not until the launch of {\it Herschel} \citep{pilbratt2010} that the $J=1$--0 transition at 243.244~$\mu$m (1232.4763~GHz) could (and would) be routinely observed \citep[e.g.,][]{neufeld2010hf,sonnentrucker2010,phillips2010,monje2011,emprechtinger2012}.  Using H$_2$ column densities estimated from other tracers such as $^{13}$CO and CH, these studies found relative abundances of $N({\rm HF})/N({\rm H}_2)$ in the (1--2)$\times10^{-8}$ range in diffuse molecular clouds, with an average value of $1.4\times10^{-8}$.

Observations of neutral atomic fluorine have also been made in several diffuse cloud sight lines \citep{federman2005,snow2007} targeting transitions at 951~\AA\ and 954~\AA\ using the {\it Far Ultraviolet Spectroscopic Explorer} ({\it FUSE}).  All sight lines where F is detected have low molecular hydrogen fractions---$f_{\rm H_2}\leq0.3$, where $f_{\rm H_2}\equiv 2N({\rm H}_2)/[N({\rm H})+2N({\rm H}_2)]$---so that F is expected to be the dominant form of fluorine.  The average fractional abundance of atomic fluorine with respect to total hydrogen---$x({\rm F})\equiv N({\rm F})/N_{\rm H}$, where $N_{\rm H}\equiv N({\rm H})+2N({\rm H}_2)$---is found to be $x({\rm F})=1.8\times10^{-8}$, about 60\%\ the solar system abundance of $2.9\times10^{-8}$ found in meteorites \citep[][and references therein]{lodders2003}.  

Taking $x({\rm F})=1.8\times10^{-8}$ to be the interstellar gas phase abundance of fluorine and assuming conditions where both fluorine and hydrogen are predominantly in molecular form results in the prediction $N({\rm HF})/N({\rm H}_2)=3.6\times10^{-8}$.  This is about twice as large as the average value determined from {\it Herschel} observations in diffuse molecular clouds, and about 100 times as large as the values found in dense molecular clouds \citep{neufeld1997,phillips2010,emprechtinger2012}.  The severe underabundance of gas phase HF in dense clouds is thought to be the result of nearly all HF freezing out onto grains in such cold environments \citep{neufeld2005,neufeld2009}, and it is possible that a similar mechanism, albeit less efficient, operates in diffuse clouds as well, resulting in the factor of 2 underabundance in those regions.  Still, all current estimates of the HF/H$_2$ ratio rely on abundance relationships between H$_2$ and some other molecule (e.g., CH, CO), and observations of that proxy molecule.

In this paper, we present the first direct determination of the HF/H$_2$ ratio by observing the $v=1$--0, $R(0)$ ro-vibrational transition of HF in sight lines where the H$_2$ column densitiy is known either from observations of electronic transitions in the ultraviolet \citep{savage1977,rachford2002,rachford2009} or of ro-vibrational quadrupole transitions in the infrared \citep[e.g.,][]{lacy1994,kulesa2002}.  While higher lying ro-vibrational transitions of HF have been observed in atmospheres of cool stars \citep[e.g.,][]{jorissen1992,cunha2003,uttenthaler2008}, this is the first detection of interstellar HF via ro-vibrational spectroscopy.

\section{OBSERVATIONS \& DATA REDUCTION} \label{section_obs_rx}

All observations were carried out using the Cryogenic High-resolution Infrared Echelle Spectrograph \citep[CRIRES;][]{kaufl2004} on UT1 at the Very Large Telescope (VLT).  Observations were performed in service mode, and CRIRES was used with its 0\farcs2 slit to provide a resolving power (resolution) of about 100,000 (3~km~s$^{-1}$), and a reference wavelength of 2502.8~nm to position the $v=1$--0, $R(0)$ ($\lambda=2.499385$~$\mu$m) and $R(1)$ ($\lambda=2.475876$~$\mu$m) transitions of HF on detectors 3 and 1, respectively.  The adaptive optics system was utilized in all cases (excluding Elias~29 and AFGL~2136~IRS~1 where no bright natural guide stars are available) to maximize starlight passing through the narrow slit.  Spectra were obtained in an ABBA pattern with 10\arcsec\ between the two nod positions and $\pm$3\arcsec\ jitter width.  Details regarding integration times and dates of observations are presented in Table \ref{tbl_obs}.  Our observing program was designed with the intent of removing atmospheric absorption features via division by model atmospheric spectra \citep{seifahrt2010}.  As such, no telluric standard stars were observed (with the exception of HR~6508, paired with the final observation of HD~154368 on 2012 Aug 02). 

Raw data images were processed using the CRIRES pipeline version 2.2.1.  Calibration techniques applied to science images include subtraction of dark frames, division by flat fields, interpolation over bad pixels, and correction for detector non-linearity effects.  Consecutive A and B nod position images were subtracted from each other to remove sky emission features.  All of the images from each nod position were combined to create average A and B images.  One-dimensional spectra were extracted from these images using the {\it apall} routine in \textsc{iraf}\footnote{http://iraf.noao.edu/}, and then imported to Igor Pro\footnote{http://www.wavemetrics.com/}.
Wavelength calibration was performed using atmospheric absorption lines, and is accurate to $\pm1$~km~s$^{-1}$.  Spectra from the A and B nod positions were averaged on a common wavelength scale.\footnote{Combination of A and B spectra is done after wavelength calibration due to a slight (about one-half pixel) shift in wavelength along the dispersion direction between the two nod positions.}

In order to remove atmospheric absorption features from science spectra, we attempted division by atmospheric model spectra.  However, simultaneous removal of both strong and weak H$_2$O features proved exceptionally difficult, such that the targeted continuum level signal-to-noise ratio (S/N) of 1000 could not be achieved.  We then attempted dividing science spectra by other science spectra with the requirement that the expected interstellar feature of the ``standard'' star be shifted at least 4~km~s$^{-1}$ with respect to the interstellar absorption feature of the science target.  This division procedure allows for interactive stretching and shifting of the standard star spectrum in the wavelength axis, and scaling of the standard star intensity according to Beer's law, and is described in \citet{mccall2001}.  Removal of both weak atmospheric lines and the wings of strong atmospheric lines using this method was found to be satisfactory for observations toward $\chi$~Oph, HD~149404, HD~152236, HD~154368 (on all nights), and Elias~29.  For AFGL~2136~IRS~1 division by a model atmospheric spectrum produced better results.  For $\zeta$~Oph the interstellar HF feature is blended with atmospheric HF absorption, and for HD~179406 the interstellar HF feature is blueshifted into the atmospheric $\nu_1$~9$_{6,3}$--8$_{5,4}$ H$_2$O line, such that division by neither a standard star nor an atmospheric model is able to adequately remove atmospheric features.  As a result, we omit these sight lines from our analysis.  After division by a standard star or model, any remaining large scale continuum fluctuations were removed by a custom-written procedure.\footnote{http://fermi.uchicago.edu/freeware/IgorPlugins/}
%Resulting spectra are plotted in Figure \ref{fig_R0spectra}.

%\section{RESULTS} \label{section_res}

\begin{figure}
\epsscale{0.5}
\plotone{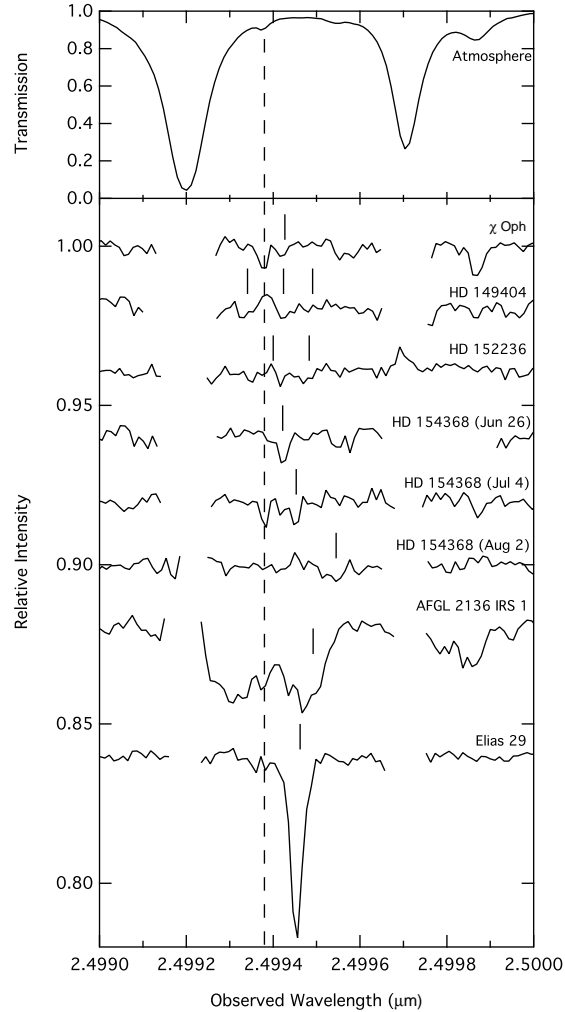}
\caption{Spectra toward target sight lines covering the $v$=1--0, $R(0)$ transition of HF. {\bf Top panel:} Observation of $\zeta$~Oph, used to show atmospheric transmission spectrum in the region of interest.  The two strongest atmospheric lines are due to the $\nu_1$~9$_{6,3}$--8$_{5,4}$ and $\nu_1$~9$_{6,4}$--8$_{5,3}$ transitions of H$_2$O at 2.499199~$\mu$m and 2.499706~$\mu$m, respectively.  HF is clearly present in the Earth's atmopshere, and the rest wavelength of the $R(0)$ transition at 2.499385~$\mu$m is marked by the vertical dashed line.  {\bf Bottom panel:} Spectra toward science targets.  Gaps in spectra are regions where division by a standard star was unable to adequately remove atmospheric absorption lines.  Solid vertical lines above spectra mark the expected position of HF absorption given the Earth's motion and previously reported interstellar gas velocities \citep[note that HD~149404 and HD~152236 have multiple velocity components reported in CH;][]{gredel1993}.  Interstellar HF is detected toward Elias~29 and AFGL~2136~IRS~1, and marginally detected toward HD~154368.  The absorption feature in the AFGL~2136~IRS~1 spectrum centered at about 2.4993~$\mu$m is from the $\nu_1$~9$_{6,3}$--8$_{5,4}$ transition of H$_2$O and likely arises in gas very close to the protostar (Indriolo et al. 2013, in preparation).
}
\label{fig_R0spectra}
\end{figure}

\begin{figure}
\epsscale{0.5}
\plotone{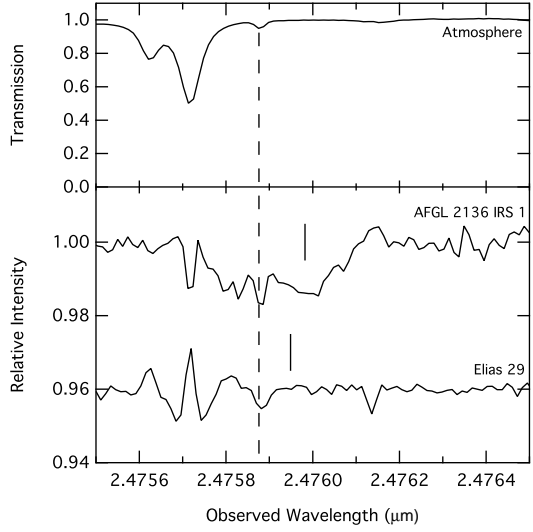}
\caption{Spectra toward target dense cloud sight lines covering the $v$=1--0, $R(1)$ transition of HF. {\bf Top panel:} Observation of $\zeta$~Oph, used to show atmospheric transmission spectrum in the region of interest.  The rest wavelength of the $R(1)$ transition at 2.475876~$\mu$m is marked by the vertical dashed line. {\bf Bottom panel:} Spectra toward AFGL~2136~IRS~1 and Elias~29. Solid vertical lines above spectra mark the expected position of HF absorption given the Earth's motion and previously reported interstellar gas velocities.  The $R(1)$ line is detected toward AFGL~2136~IRS~1, but not toward Elias~29.  As before, the absorption feature to the left of the HF line toward AFGL~2136~IRS~1 is due to water near the protostar itself, in this case, the $\nu_1$~10$_{7,4}$--9$_{6,3}$ transition.}
\label{fig_R1spectra}
\end{figure}

\begin{figure}
\epsscale{0.5}
\plotone{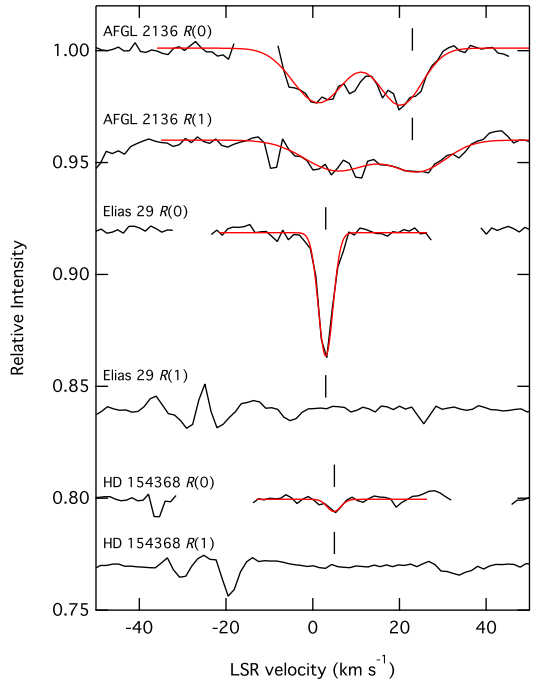}
\caption{Spectra toward AFGL~2136~IRS~1, Elias~29, and HD~154368 showing the HF $R(0)$ and $R(1)$ lines in velocity space.  The $R(0)$ and $R(1)$ absorption features toward AFGL~2136~IRS~1 are at roughly the same velocity, but have different profiles and line-widths (see Table \ref{tbl_absorption}).  Only the $R(0)$ line is detected toward Elias~29 and HD~154368.  Features between $-15$~km~s$^{-1}$ and $-40$~km~s$^{-1}$ are due to imperfect removal of an atmospheric water line.  For HD~154368, these spectra are the combination of all three nights of data.  Red curves show Gaussian fits to the HF (and H$_2$O for AFGL~2136~IRS~1) absorption features.}
\label{fig_velspectra}
\end{figure}

Spectra covering the $R(0)$ transition resulting from our reduction procedures are shown in Figure \ref{fig_R0spectra}.  $\chi$~Oph, HD~149404, and HD~152236 show no evidence of interstellar HF absorption, while HF is clearly detected toward Elias~29 and AFGL~2136~IRS~1\footnote{The HF $R(0)$ line toward AFGL~2136~IRS~1 is partially blended with a water feature that is also astrophysical in origin.}.  Spectra of HD~154368 from all three nights when observations were taken show marginal absorption features at the positions expected given the Earth's motion and known interstellar gas velocities.  Because absorption is seen at the expected wavelength on all three nights, we are more confident in considering this a detection.  Figure \ref{fig_R1spectra} shows spectra of the two dense cloud sight lines covering the $R(1)$ transition.  Although no $R(1)$ feature is seen toward Elias~29, HF absorption out of the $J=1$ level is detected toward AFGL~2136~IRS~1\footnote{Again, partially blended with an astrophysical water line.}.  Figure \ref{fig_velspectra} shows spectra in velocity space covering both the $R(0)$ and $R(1)$ transitions in all three sight lines where the $R(0)$ line is detected.  Interestingly, the $R(0)$ and $R(1)$ line widths and line profiles differ toward AFGL~2136~IRS~1, possibly indicating different physical conditions (e.g., temperature, density) in the gas from which the absorption arises, as will be discussed below.

\section{ANALYSIS}

HF absorption features were fit with Gaussian functions as described in \citet{indriolo2012} for the purpose of determining equivalent widths, velocity FWHM, and interstellar gas velocities.  These parameters are reported in Table \ref{tbl_absorption}, along with HF column densities derived assuming optically thin absorption.  In the case of HD~154368, spectra from all three nights were shifted to the LSR frame and averaged prior to this fitting procedure.  For AFGL~2136~IRS~1 the HF $R(0)$ and H$_2$O $\nu_1$~9$_{6,3}$--8$_{5,4}$ lines were fit simultaneously, as were the HF $R(1)$ and H$_2$O $\nu_1$~10$_{7,4}$--9$_{6,3}$ lines.  For non-detections, uncertainties in the equivalent width were derived from the standard deviation on the continuum where the absorption lines were expected, the wavelength step between pixels, and the number of pixels expected to contain absorption assuming a line with FWHM=3~km~s$^{-1}$ (for Elias~29 and HD~154368 the measured FWHM of the $R(0)$ lines were used in determining uncertainties in the equivalent width of the $R(1)$ line).  These $1\sigma$ uncertainties in equivalent width and column density are also reported in Table \ref{tbl_absorption}.

Upper limits on $N({\rm HF})$ in a single cloud component are taken to be 3 times the uncertainty of the HF column densities for non-detections.  In sight lines where multiple velocity components are observed in CH \citep[2 components toward HD~152236 and 3 components toward HD~149404;][]{gredel1993}, total line-of-sight upper limits on $N({\rm HF})$ are taken to be the upper limit determined for a single cloud component multiplied by the number of observed components.  These values are reported in Table \ref{tbl_ratios}, along with H$_2$ and CH column densities from the literature for sight lines we observed in HF, and the relative abundance of HF with respect to these species.  Figure \ref{fig_columns} shows $N({\rm HF})$ versus $N({\rm H}_2)$ (left panel) for the six sight lines in Table \ref{tbl_ratios}, and $N({\rm HF})$ versus $N({\rm CH})$ (right panel) for sight lines in this study and reported in the literature.  Also shown in Figure \ref{fig_columns} are the expected abundance relationships---$N({\rm HF})/N({\rm H}_2)=3.6\times10^{-8}$ and $N({\rm HF})/N({\rm CH})=1$---assuming conditions where nearly all hydrogen is H$_2$, nearly all gas-phase fluorine is HF, and $N({\rm CH})/N({\rm H}_2)=3.5\times10^{-8}$ \citep{sheffer2008}, as well as abundances predicted by chemical models presented in \citet{neufeld2005}.

\section{DISCUSSION}

\begin{figure}
\epsscale{1.0}
\plotone{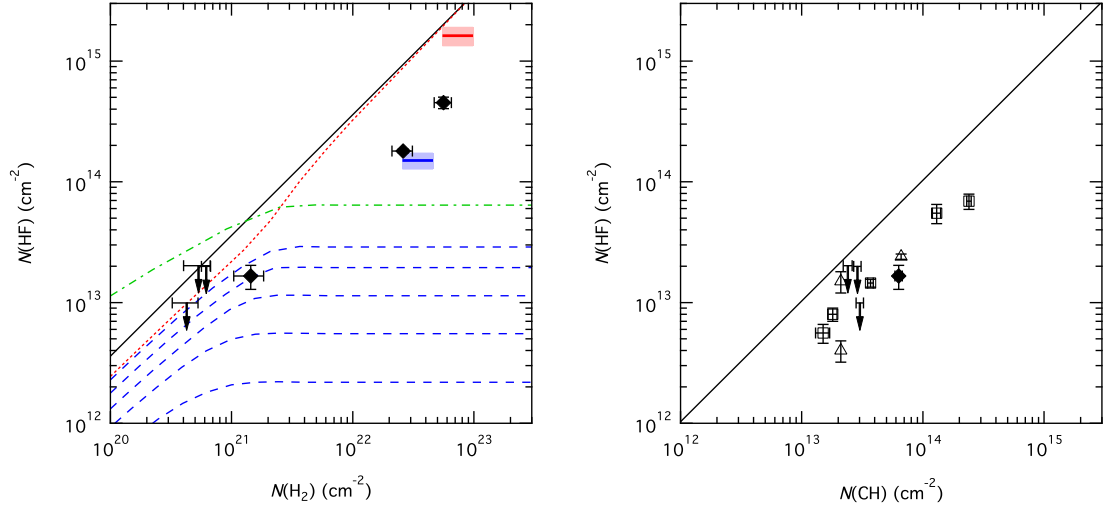}
\caption{Plots showing column densities of HF, CH, and H$_2$. {\bf Left panel:} $N({\rm HF})$ vs. $N({\rm H}_2)$ for the six sight lines reported in this study.  Filled black diamonds mark detections of HF, and black arrows mark upper limits to $N({\rm HF})$ reported in Table \ref{tbl_ratios}.  The red and blue shaded regions show a decomposition of the AFGL~2136~IRS~1 data into a warm and cold component, respectively (as described in Section \ref{subsec_2136}), where dark lines mark the derived HF column densities, extending horizontally to show the range of inferred H$_2$ column densities, and the lighter shaded regions extend vertically to mark $1\sigma$ uncertainties in the HF column densities.  The solid line shows HF/H$_2=3.6\times10^{-8}$.  Dashed (blue), dotted (red), and dash-dot (green) curves show results from the models of \citet{neufeld2005} with different densities ($n_{\rm H}$), ultraviolet radiation fields ($\chi_{\rm UV}$), and fluorine depletion schemes.  The red dotted line is for a model cloud with F depletion independent of $A_V$, $\chi_{\rm UV}=1$, and $n_{\rm H}=10^2$~cm$^{-3}$ (see their Figure 7).  Blue dashed curves show model results for F depletion variable with $A_V$, $\chi_{\rm UV}=1$, and $n_{\rm H}=10^2$, $10^{2.5}$, $10^{3}$, $10^{3.5}$,  and $10^{4}$~cm$^{-3}$ from top to bottom (see their Figure 9).  The green dash-dot curve is for F depletion variable with $A_V$, $\chi_{\rm UV}=1000$, and $n_{\rm H}=10^4$~cm$^{-3}$ (also their Figure 9).  {\bf Right panel:} $N({\rm HF})$ vs. $N({\rm CH})$ for sight lines reported in this study, as well as sight lines where both HF and CH have been observed with {\it Herschel}.  The filled black diamonds and arrows as are before, while the open black squares mark HF and CH column densities reported in \citet{sonnentrucker2010} and \citet{gerin2010ch}, and open black triangles are from \citet{emprechtinger2012}.  The solid line shows the $\sim$ 1:1 ratio of HF to CH expected given fully molecular fluorine and the CH/H$_2$ ratio found by \citet{sheffer2008}.  Elias~29 and AFGL~2136~IRS~1 are not in this panel as no CH column densities are reported in the literature.}
\label{fig_columns}
\end{figure}

In the left panel of Figure \ref{fig_columns} it is clear that our measured HF abundances do not match the expected abundances for fully molecular H$_2$ and HF.  Aside from the upper limits toward HD~149404 and HD~152236, which are close to the predicted ratio, all measured HF abundances are lower than expected.  The right-hand panel of Figure \ref{fig_columns} shows that the measured HF abundances also do not match the expected 1:1 ratio with CH, again with observed values below predicted values.  Measured CH abundance are consistent with the relationship $N({\rm CH})/N({\rm H}_2)=3.5_{-1.4}^{+2.1}\times10^{-8}$ found by \citet{sheffer2008}, suggesting that we are seeing low abundances of HF, rather than high abundances of CH.  Our findings agree with HF/CH ratios determined from {\it Herschel} observations of both species \citep{sonnentrucker2010,gerin2010ch,emprechtinger2012}.  As such, it seems likely that the fully molecular fluorine assumption is incorrect, and that one or more mechanisms must decrease the formation rate of HF, increase the destruction rate of HF, or remove HF from the gas phase.

%\citet{neufeld2005} presented a simple model of fluorine chemistry for a plane-parallel cloud illuminated on both sides.  In addition to studying the pure gas-phase reaction network, they also investigated the effects of F and HF depletion onto grains.  Select results from their parameter study (varied UV field, varied density, depletion (in)variable with $A_V$) are plotted in the left-hand panel of Figure \ref{fig_columns}.  Unlike the fully molecular HF and H$_2$ assumption, the model results of \citet{neufeld2005} can reproduce observed HF/H$_2$ ratios.  The model with $\chi_{\rm UV}=1$, $n_{\rm H}=10^{2.5}$~cm$^{-3}$, and variable depletion included does an excellent job of predicting $N({\rm HF})/N({\rm H}_2)$ toward HD~154368, which has an estimated gas density of $n_{\rm H}=240$~cm$^{-3}$ \citep{sonnentrucker2007}.  However, models with depletion are unable to reproduce our findings toward the dense cloud sight lines Elias~29 and AFGL~2136~IRS~1, drastically underpredicting observed HF column densities even for high ($\chi_{\rm UV}=1000$) UV fields.

\citet{neufeld2005} discussed interstellar fluorine chemistry in detail, and presented a simple model for a plane-parallel geometry cloud illuminated on both sides by the interstellar radiation field.  In addition to studying the pure gas-phase reaction network, they also investigated the effects of F and HF depletion onto grains.  Select results from their study 
%(varied UV field\footnote{UV flux is given in terms of the ``standard'' UV background from \citet{draine1978}, denoted as $\chi_{\rm UV}=1$.}, varied density, depletion (in)variable with $A_V$) 
are plotted in the left-hand panel of Figure \ref{fig_columns}.  The red dotted curve is for a model cloud with F depletion independent of $A_V$, $\chi_{\rm UV}=1$,\footnote{UV flux is given in terms of the ``standard'' UV background from \citet{draine1978}, denoted as $\chi_{\rm UV}=1$.} and $n_{\rm H}=10^2$~cm$^{-3}$.  While the model is consistent with our upper limits, it cannot reproduce the observed HF/H$_2$ ratios in sight lines where HF is detected.  Increasing the density does not improve the agreement, as model results shift toward the HF/H$_2=3.6\times10^{-8}$ fully molecular scenario \citep[see Figure 7 in][]{neufeld2005}.  Similarly, increasing the impinging UV field only serves to worsen the agreement with our observations.  The purely gas-phase chemical models still predict that most fluorine is in the form of HF over a wide range of input conditions, suggesting that formation and/or destruction rates of HF do not vary enough to account for the observed abundances.

Model results with variable fluorine depletion are also shown in Figure \ref{fig_columns}, and in some cases do a better job of reproducing observations.  Blue dashed curves show model results for F depletion variable with $A_V$, $\chi_{\rm UV}=1$, and $n_{\rm H}=10^2$, $10^{2.5}$, $10^{3}$, $10^{3.5}$,  and $10^{4}$~cm$^{-3}$ (from top to bottom).  \citet{sonnentrucker2007} estimated a density of $n_{\rm H}=240$~cm$^{-3}$ from C$_2$ observations toward HD~154368, and the model curve with  $n_{\rm H}=10^{2.5}$~cm$^{-3}$ is in excellent agreement with our observed HF column density.  However, models with depletion are unable to reproduce our findings toward the dense cloud sight lines Elias~29 and AFGL~2136~IRS~1, drastically underpredicting observed HF column densities even with an increased UV field.\footnote{The green dash-dot curve is for a model with F depletion variable with $A_V$, $\chi_{\rm UV}=1000$, and $n_{\rm H}=10^4$~cm$^{-3}$.} To determine why this is so, we first explore properties of the Elias~29 and AFGL~2136~IRS~1 sight lines in detail.

\subsection{Elias~29}

The line of sight toward the low-mass protostar Elias~29 passes through several different environments.  Multiple velocity components are traced by emission in CO, $^{13}$CO, C$^{18}$O, CS, C$^{34}$S, HCO$^+$, H$^{13}$CO$^+$, H$_2$CO, and CH$_3$OH \citep{boogert2002}.  Two foreground clouds at $v_{\rm LSR}=2.7$~km~s$^{-1}$ and 3.8~km~s$^{-1}$ are thought to be unassociated with the protostar, while some combination of the molecular ridge in which the protostar is embedded and the protostellar envelope itself is thought to give rise to emission and self-absorption seen at  $v_{\rm LSR}=5$~km~s$^{-1}$.  Densities and kinetic temperatures of $n_{\rm H}\sim10^4$~cm$^{-3}$ and $T\sim20$~K in the foreground clouds and $n_{\rm H}\sim4\times10^5$~cm$^{-3}$ and $T\sim15$~K in the molecular ridge have been estimated from observations of C$^{18}$O and HCO$^+$ \citep{boogert2002}.  Warm ($T\sim500$~K) CO and H$_2$O are seen via ro-vibrational transitions in absorption (at low spectral resolution, thus lacking velocity information), likely arising from a hot core or flared disk \citep{boogert2000}.  This warm gas is also traced by emission arising from vibrationally excited states of $^{12}$CO and $^{13}$CO seen in high spectral resolution observations near 4.6~$\mu$m \citep{herczeg2011}.  Additionally, absorption in $^{12}$CO from 0 to -90~km~s$^{-1}$ (with respect to systemic) is thought to trace a wind/outflow component.

The total H$_2$ column density in all three cold components---inferred from C$^{18}$O $J=2$--1 emission---is $N({\rm H}_2)=2.9\times10^{22}$~cm$^{-2}$, in good agreement with $(2.6\pm0.5)\times10^{22}$~cm$^{-2}$ determined directly from observations of the $v=1$--0, $S(0)$ transition of H$_2$ \citep{kulesa2002}.  This suggests that the H$_2$ column density listed in Table \ref{tbl_ratios} and used in Figure \ref{fig_columns} is the total line-of-sight H$_2$ column density, and does not discern between the different cloud components.  However, the observed HF absorption feature is relatively narrow (FWHM=4.0~km~s$^{-1}$) and has a velocity centroid that favors association with one or both of the foreground clouds.  Given the small spread in velocity along this sight line though, it is difficult to ascertain which cloud(s) may give rise to HF absorption.  It is quite possible that we have underestimated the HF/H$_2$ ratio by including H$_2$ that is not associated with the component where HF is observed (and consequently overestimated the ratio in components where HF absorption does not arise).  Making this assumption, only in the case where all HF is associated with the 2.7~km~s$^{-1}$ cloud \citep[$N({\rm H}_2)\sim5\times10^{21}$~cm$^{-2}$;][]{boogert2002} can the models reproduce $N({\rm HF})$ and $N({\rm H}_2)$, and even then, it is only models with constant depletion that are capable of doing so.

\subsection{AFGL~2136~IRS~1} \label{subsec_2136}

AFGL~2136~IRS~1 is an embedded massive protostar that shows a complex 3-lobed morphology at 2.2~$\mu$m---thought to trace the cavity walls of a bipolar outflow---and a polarization signature indicative of a nearly edge-on disk \citep{kastner1992,murakawa2008}.  
Emission maps in $^{12}$CO $J=1$--0 and $J=2$--1 and $^{13}$CO $J=1$--0 and $J=2$--1 show massive molecular outflows that support this picture, and give a systemic velocity of $v_{\rm LSR}=22$~km~s$^{-1}$ \citep{kastner1994}.  Absorption due to the $v=1$--0 and $v=2$--1 bands of $^{12}$CO and the $v=1$--0 band of $^{13}$CO reveal both cold ($T\sim20$~K) and warm ($T\sim580$~K) components at $v_{\rm LSR}=26.5\pm2.8$~km~s$^{-1}$ \citep{mitchell1990co}.
%, likely arising from gas located close to the protostar.  
%Warm ($T\sim500$~K) gas is also observed in absorption from the $\nu_2$ ro-vibrational band of H$_2$O \citep{boonman2003}, and our own observations targeting HF have revealed absorption in the $\nu_1$ and $\nu_3$ bands of H$_2$O arising from levels as high as $J_{K_{a},K_{c}}=17_{0,17}$, more than 4000~K above the ground state (Indriolo 2013, in preparation).
Warm ($T\sim500$~K) gas is also observed in absorption from the $\nu_2$ ro-vibrational band of H$_2$O \citep{boonman2003}, and our own observations targeting HF have revealed absorption in the $\nu_1$ and $\nu_3$ bands of H$_2$O arising from more than 30 different rotational levels with relative populations consistent with a temperature of $500\pm20$~K (Indriolo 2013, in preparation).

As mentioned above, a very high density ($n_{\rm H}\gtrsim10^9$) and/or exceptional radiative pumping is necessary to maintain substantial population in the $J=1$ level of HF, making our detection of the $R(1)$ line of HF toward AFGL~2136~IRS~1 rather surprising.  However,  studies of CO \citep{mitchell1990co} and H$_2$O \citep{boonman2003} suggest that the absorbing gas toward AFGL~2136~IRS~1 is dominated by warm ($T\sim500$~K), dense ($n_{\rm H}>10^{10}$~cm$^{-3}$) gas in close proximity to the central protostar, perhaps located in a protostellar disk.  In this case, relative populations in rotationally excited levels of HF are expected to be thermalized, and the $R(1)$ line is detectable.  The profile of the HF $R(1)$ line is in good agreement with those of most excited water lines (Indriolo et al. 2013, in preparation), suggesting that the two species reside in similar gas conditions.  The HF $R(0)$ line, however, is narrower than the $R(1)$ line and most water lines, implying a different origin.

In thermal equilibrium at $T=500$~K the relative population in the $J=1$ and $J=0$ levels is expected to be $N(1)/N(0)=2.7$, much larger than the observed line-of-sight ratio of 1.03.  However, this can be reconciled if we ascribe a portion of the $R(0)$ absorption to cold, moderately dense %($n_{\rm H}\sim10^6$~cm$^{-3}$) 
gas in the protostellar envelope.  Assuming that HF is in thermal equilibrium in the warm gas results in $N(0)_{warm}=0.92\times10^{14}$~cm$^{-2}$ and $N(0)_{cold}=1.50\times10^{14}$~cm$^{-2}$.  Under this assumption levels with $J\geq2$ will also be populated, such that $N({\rm HF})_{warm}/N(1)=6.5$, and the total HF column density in warm gas is $N({\rm HF})_{warm}=1.62\times10^{15}$~cm$^{-2}$.  As all HF in cold gas is still expected to be in the $J=0$ level, $N({\rm HF})_{cold}=1.50\times10^{14}$~cm$^{-2}$.  

Determining HF/H$_2$ ratios then requires finding the H$_2$ columns in both the warm and cold components.  To do so, we set the temperature in the warm gas to 500~K, and allow the temperature in the cold gas to vary from 20--80~K,\footnote{The 20~K limit is based on the cold component found by \citet{mitchell1990co} in $^{13}$CO and the 80~K limit on an H$_2$ {\it ortho}-to-{\it para} ratio of 1.} calculating the fractional population of H$_2$ expected in the $J=0$ level for all cases.  The sum $N({\rm H_2};J=0)_{warm}+N({\rm H_2};J=0)_{cold}$ is constrained to equal $3\times10^{22}$~cm$^{-2}$ \citep[total line-of-sight column density in the $J=0$ level of H$_2$ reported by][]{kulesa2002}, while the ratio of warm H$_2$ to cold H$_2$ is assumed to be the same as that found by \citet{mitchell1990co} for warm to cold $^{13}$CO: 2.14.  Given the 20--80~K range assumed for the temperature of the cold gas, the total H$_2$ column density in the warm component is $(5.5$--$9.8)\times10^{22}$~cm$^{-2}$, and the total H$_2$ column density in the cold component is $(2.6$--$4.6)\times10^{22}$~cm$^{-2}$.  Taking the HF column densities from above, we find HF/H$_2$ ratios of $(1.7$--$2.9)\times10^{-8}$ for the warm gas, and $(0.33$--$0.58)\times10^{-8}$ for the cold gas.\footnote{Note that a larger HF/H$_2$ ratio corresponds to a lower temperature in the cold component.}

The HF and H$_2$ column densities found from this analysis are shown in Figure \ref{fig_columns}.  Abundances in the warm gas component are nearly consistent with the simple scenario where all fluorine is in the form of HF.  Given that H$_2$O is observed to be predominantly in the gas phase in this region, it is likely that depletion of fluorine onto grains is negligible (due to the grains being warm), and the simple chemistry mentioned in Section \ref{section_intro} applies.  Relative abundances of HF and H$_2$ in the cold gas component, however, remain inconsistent with any of the models discussed herein.

\subsection{Discrepancy between models and observations}

The chemical models of \citet{neufeld2005} that invoke fluorine depletion varying with visual extinction assume the rate at which F and HF stick to grains is proportional to the dust density (which is proportional to the density of H nuclei assuming a fixed gas-to-dust ratio).  HF is only removed from grains via UV photodesorption, a process inhibited in cloud interiors where UV photons are only emitted following cosmic-ray induced excitation of H$_2$ Lyman and Werner bands.  As a result, the models predict that all HF inside of dense clouds should be frozen onto grains.  However, dust grains in the inner portions of a protostellar envelope are expected to be warmed by the central radiation source---an effect not included in the depletion model---leading to the evaporation of icy grain mantles.  We have already alluded to this likely being the scenario for the warm gas component toward AFGL~2136~IRS~1.  It is possible that some portion of the HF absorption seen toward Elias~29 also arises from warm gas close to the protostar, but the lack of $R(1)$ absorption means we cannot attempt the same warm/cold decomposition performed for AFGL~2136~IRS~1 above.  However, this mechanism still cannot explain the larger than expected HF/H$_2$ ratio in the cold component of AFGL~2136~IRS~1 unless the grain temperature is significantly warmer than the gas temperature.

It is possible that some amount of the differences between observed and model abundances stems from geometric effects.  Model results presented in Figure \ref{fig_columns} are for plane-parallel clouds viewed face-on.  For inclined viewing angles, sight lines sample more material near the edge of the cloud, where, in the case of fluorine depletion increasing with $A_V$, HF has not yet frozen onto grains.  As a result, the line-of-sight HF/H$_2$ ratio increases for inclined viewing angles.  A similar effect occurs for the same reason if we consider multiple clouds along the line-of-sight.  Still, these geometric effects alone cannot bring the model curve for a dense cloud ($\chi_{\rm UV}=1$, $n_{\rm H}=10^4$~cm$^{-3}$; bottom dashed blue curve in the left panel of Figure \ref{fig_columns}) into agreement with observations.

The constant densities assumed in these cloud models may also contribute to discrepancies with observations.  For the $n_{\rm H}=10^4$~cm$^{-3}$ model, HF is efficiently removed from the gas phase even at the edge of the cloud because the sticking coefficient is proportional to the gas density.  Dense clouds are expected to be surrounded by diffuse material though, where HF will remain in the gas phase.  A line of sight probing a dense cloud with a diffuse envelope would contain an HF abundance dependent almost entirely on the size of the envelope, because little or no fluorine depletion occurs there.  In this way, a cloud that consists predominantly of dense gas can still show an appreciable HF column density if it is surrounded by diffuse material.  

%Most likely, the model densities, geometric effects, and the lack of HF evaporation at high temperatures all contribute to discrepancies between observed and modeled abundances, but it is difficult to ascertain which may have the greatest effect.

%It is interesting that even after attempting to account for multiple cloud components with different physical conditions along both dense cloud sight lines, some of our inferred HF/H$_2$ ratios cannot be predicted by model results.  Instead, they fall between model results assuming a constant fluorine depletion level and a fluorine depletion that increases with visual extinction.  Some amount of increased depletion is necessary to explain the observed HF/H$_2$ ratios, but perhaps the models overpredict depletion at large $A_{\rm V}$. 

\section{SUMMARY} \label{section_sum}

For the first time, we have detected the $v=1$--0, $R(0)$ transition of HF in interstellar gas.  Additionally, we have detected both the $R(0)$ and $R(1)$ transitions in the inner envelope or protostellar disk around AFGL~2136~IRS~1.  This warm gas associated with a massive protostar has an HF/H$_2$ ratio similar to that predicted by the scenario where all hydrogen is in molcular form and all fluorine is in HF, while observed HF abundances in interstellar gas are much lower.  Chemical models assuming increasing depletion of HF onto grains with increasing $A_{\rm V}$ can reproduce the HF/H$_2$ ratio observed toward HD~154368, but not in the cold, dense component toward AFGL~2136~IRS~1 and Elias~29.  It is possible that some assumptions used in modeling the fluorine chemistry---constant densities, plane-parallel geometry, the lack of HF evaporation at high temperatures---all contribute to discrepancies between observed and modeled abundances, but it is difficult to ascertain which may have the greatest effect.  More detailed chemical models and an expanded sample of HF observations may help to elucidate the most important factors. 

\mbox{}
The authors would like to thank the anonymous referee for comments and suggestions.  This research was performed, in part, through a JPL contract funded by the National Aeronautics and Space Administration.  M.J.R. is supported by NASA Collaborative Agreement NNX10ac27a.

%\bibliographystyle{apj}
%\bibliography{indy_master}

\begin{thebibliography}{41}
\expandafter\ifx\csname natexlab\endcsname\relax\def\natexlab#1{#1}\fi

\bibitem[{{Boogert} {et~al.}(2002){Boogert}, {Hogerheijde}, {Ceccarelli},
  {Tielens}, {van Dishoeck}, {Blake}, {Latter}, \& {Motte}}]{boogert2002}
{Boogert}, A.~C.~A., {Hogerheijde}, M.~R., {Ceccarelli}, C., {et~al.} 2002,
  \apj, 570, 708

\bibitem[{{Boogert} {et~al.}(2000){Boogert}, {Tielens}, {Ceccarelli},
  {Boonman}, {van Dishoeck}, {Keane}, {Whittet}, \& {de Graauw}}]{boogert2000}
{Boogert}, A.~C.~A., {Tielens}, A.~G.~G.~M., {Ceccarelli}, C., {et~al.} 2000,
  \aap, 360, 683

\bibitem[{{Boonman} \& {van Dishoeck}(2003)}]{boonman2003}
{Boonman}, A.~M.~S., \& {van Dishoeck}, E.~F. 2003, \aap, 403, 1003

\bibitem[{{Cunha} {et~al.}(2003){Cunha}, {Smith}, {Lambert}, \&
  {Hinkle}}]{cunha2003}
{Cunha}, K., {Smith}, V.~V., {Lambert}, D.~L., \& {Hinkle}, K.~H. 2003, \aj,
  126, 1305

\bibitem[{{Draine}(1978)}]{draine1978}
{Draine}, B.~T. 1978, \apjs, 36, 595

\bibitem[{{Emprechtinger} {et~al.}(2012){Emprechtinger}, {Monje}, {van der
  Tak}, {van der Wiel}, {Lis}, {Neufeld}, {Phillips}, \&
  {Ceccarelli}}]{emprechtinger2012}
{Emprechtinger}, M., {Monje}, R.~R., {van der Tak}, F.~F.~S., {et~al.} 2012,
  \apjs, 756, 136

\bibitem[{{Federman} {et~al.}(2005){Federman}, {Sheffer}, {Lambert}, \&
  {Smith}}]{federman2005}
{Federman}, S.~R., {Sheffer}, Y., {Lambert}, D.~L., \& {Smith}, V.~V. 2005,
  \apj, 619, 884

\bibitem[{{Gerin} {et~al.}(2010){Gerin}, {de Luca}, {Goicoechea}, {Herbst},
  {Falgarone}, {Godard}, {Bell}, {Coutens}, {Ka{\'z}mierczak}, {Sonnentrucker},
  {Black}, {Neufeld}, {Phillips}, {Pearson}, {Rimmer}, {Hassel}, {Lis},
  {Vastel}, {Boulanger}, {Cernicharo}, {Dartois}, {Encrenaz}, {Giesen},
  {Goldsmith}, {Gupta}, {Gry}, {Hennebelle}, {Hily-Blant}, {Joblin},
  {Ko{\l}os}, {Kre{\l}owski}, {Mart{\'{\i}}n-Pintado}, {Monje}, {Mookerjea},
  {Perault}, {Persson}, {Plume}, {Salez}, {Schmidt}, {Stutzki}, {Teyssier},
  {Yu}, {Contursi}, {Menten}, {Geballe}, {Schlemmer}, {Morris}, {Hatch},
  {Imram}, {Ward}, {Caux}, {G{\"u}sten}, {Klein}, {Roelfsema}, {Dieleman},
  {Schieder}, {Honingh}, \& {Zmuidzinas}}]{gerin2010ch}
{Gerin}, M., {de Luca}, M., {Goicoechea}, J.~R., {et~al.} 2010, \aap, 521, L16

\bibitem[{{Gredel} {et~al.}(1993){Gredel}, {van Dishoeck}, \&
  {Black}}]{gredel1993}
{Gredel}, R., {van Dishoeck}, E.~F., \& {Black}, J.~H. 1993, \aap, 269, 477

\bibitem[{{Herczeg} {et~al.}(2011){Herczeg}, {Brown}, {van Dishoeck}, \&
  {Pontoppidan}}]{herczeg2011}
{Herczeg}, G.~J., {Brown}, J.~M., {van Dishoeck}, E.~F., \& {Pontoppidan},
  K.~M. 2011, \aap, 533, A112

\bibitem[{{Indriolo} \& {McCall}(2012)}]{indriolo2012}
{Indriolo}, N., \& {McCall}, B.~J. 2012, \apj, 745, 91

\bibitem[{{Jorissen} {et~al.}(1992){Jorissen}, {Smith}, \&
  {Lambert}}]{jorissen1992}
{Jorissen}, A., {Smith}, V.~V., \& {Lambert}, D.~L. 1992, \aap, 261, 164

\bibitem[{{Kastner} {et~al.}(1992){Kastner}, {Weintraub}, \&
  {Aspin}}]{kastner1992}
{Kastner}, J.~H., {Weintraub}, D.~A., \& {Aspin}, C. 1992, \apj, 389, 357

\bibitem[{{Kastner} {et~al.}(1994){Kastner}, {Weintraub}, {Snell}, {Sandell},
  {Aspin}, {Hughes}, \& {Baas}}]{kastner1994}
{Kastner}, J.~H., {Weintraub}, D.~A., {Snell}, R.~L., {et~al.} 1994, \apj, 425,
  695

\bibitem[{{K\"{a}ufl} {et~al.}(2004){K\"{a}ufl}, {Ballester}, {Biereichel},
  {Delabre}, {Donaldson}, {Dorn}, {Fedrigo}, {Finger}, {Fischer}, {Franza},
  {Gojak}, {Huster}, {Jung}, {Lizon}, {Mehrgan}, {Meyer}, {Moorwood}, {Pirard},
  {Paufique}, {Pozna}, {Siebenmorgen}, {Silber}, {Stegmeier}, \&
  {Wegerer}}]{kaufl2004}
{K\"{a}ufl}, H., {Ballester}, P., {Biereichel}, P., {et~al.} 2004, \procspie,
  5492, 1218

\bibitem[{{Kessler} {et~al.}(1996){Kessler}, {Steinz}, {Anderegg}, {Clavel},
  {Drechsel}, {Estaria}, {Faelker}, {Riedinger}, {Robson}, {Taylor}, \&
  {Xim{\'e}nez de Ferr{\'a}n}}]{kessler1996}
{Kessler}, M.~F., {Steinz}, J.~A., {Anderegg}, M.~E., {et~al.} 1996, \aap, 315,
  L27

\bibitem[{{Kulesa}(2002)}]{kulesa2002}
{Kulesa}, C.~A. 2002, PhD thesis, The University of Arizona

\bibitem[{{Lacy} {et~al.}(1994){Lacy}, {Knacke}, {Geballe}, \&
  {Tokunaga}}]{lacy1994}
{Lacy}, J.~H., {Knacke}, R., {Geballe}, T.~R., \& {Tokunaga}, A.~T. 1994,
  \apjl, 428, L69

\bibitem[{{Lodders}(2003)}]{lodders2003}
{Lodders}, K. 2003, \apj, 591, 1220

\bibitem[{{McCall}(2001)}]{mccall2001}
{McCall}, B.~J. 2001, PhD thesis, The University of Chicago

\bibitem[{{Mitchell} {et~al.}(1990){Mitchell}, {Maillard}, {Allen}, {Beer}, \&
  {Belcourt}}]{mitchell1990co}
{Mitchell}, G.~F., {Maillard}, J.-P., {Allen}, M., {Beer}, R., \& {Belcourt},
  K. 1990, \apjs, 363, 554

\bibitem[{{Monje} {et~al.}(2011){Monje}, {Emprechtinger}, {Phillips}, {Lis},
  {Goldsmith}, {Bergin}, {Bell}, {Neufeld}, \& {Sonnentrucker}}]{monje2011}
{Monje}, R.~R., {Emprechtinger}, M., {Phillips}, T.~G., {et~al.} 2011, \apjl,
  734, L23

\bibitem[{{Murakawa} {et~al.}(2008){Murakawa}, {Preibisch}, {Kraus}, \&
  {Weigelt}}]{murakawa2008}
{Murakawa}, K., {Preibisch}, T., {Kraus}, S., \& {Weigelt}, G. 2008, \aap, 490,
  673

\bibitem[{{Neufeld} \& {Wolfire}(2009)}]{neufeld2009}
{Neufeld}, D.~A., \& {Wolfire}, M.~G. 2009, \apj, 706, 1594

\bibitem[{{Neufeld} {et~al.}(2005){Neufeld}, {Wolfire}, \&
  {Schilke}}]{neufeld2005}
{Neufeld}, D.~A., {Wolfire}, M.~G., \& {Schilke}, P. 2005, \apj, 628, 260

\bibitem[{{Neufeld} {et~al.}(1997){Neufeld}, {Zmuidzinas}, {Schilke}, \&
  {Phillips}}]{neufeld1997}
{Neufeld}, D.~A., {Zmuidzinas}, J., {Schilke}, P., \& {Phillips}, T.~G. 1997,
  \apjl, 488, L141

\bibitem[{{Neufeld} {et~al.}(2010){Neufeld}, {Sonnentrucker}, {Phillips},
  {Lis}, {de Luca}, {Goicoechea}, {Black}, {Gerin}, {Bell}, {Boulanger},
  {Cernicharo}, {Coutens}, {Dartois}, {Kazmierczak}, {Encrenaz}, {Falgarone},
  {Geballe}, {Giesen}, {Godard}, {Goldsmith}, {Gry}, {Gupta}, {Hennebelle},
  {Herbst}, {Hily-Blant}, {Joblin}, {Ko{\l}os}, {Kre{\l}owski},
  {Mart{\'{\i}}n-Pintado}, {Menten}, {Monje}, {Mookerjea}, {Pearson},
  {Perault}, {Persson}, {Plume}, {Salez}, {Schlemmer}, {Schmidt}, {Stutzki},
  {Teyssier}, {Vastel}, {Yu}, {Cais}, {Caux}, {Liseau}, {Morris}, \&
  {Planesas}}]{neufeld2010hf}
{Neufeld}, D.~A., {Sonnentrucker}, P., {Phillips}, T.~G., {et~al.} 2010, \aap,
  518, L108

\bibitem[{{Phillips} {et~al.}(2010){Phillips}, {Bergin}, {Lis}, {Neufeld},
  {Bell}, {Wang}, {Crockett}, {Emprechtinger}, {Blake}, {Caux}, {Ceccarelli},
  {Cernicharo}, {Comito}, {Daniel}, {Dubernet}, {Encrenaz}, {Gerin}, {Giesen},
  {Goicoechea}, {Goldsmith}, {Herbst}, {Joblin}, {Johnstone}, {Langer},
  {Latter}, {Lord}, {Maret}, {Martin}, {Melnick}, {Menten}, {Morris},
  {M{\"u}ller}, {Murphy}, {Ossenkopf}, {Pearson}, {P{\'e}rault}, {Plume},
  {Qin}, {Schilke}, {Schlemmer}, {Stutzki}, {Trappe}, {van der Tak}, {Vastel},
  {Yorke}, {Yu}, {Zmuidzinas}, {Boogert}, {G{\"u}sten}, {Hartogh}, {Honingh},
  {Karpov}, {Kooi}, {Krieg}, \& {Schieder}}]{phillips2010}
{Phillips}, T.~G., {Bergin}, E.~A., {Lis}, D.~C., {et~al.} 2010, \aap, 518,
  L109

\bibitem[{{Pilbratt} {et~al.}(2010){Pilbratt}, {Riedinger}, {Passvogel},
  {Crone}, {Doyle}, {Gageur}, {Heras}, {Jewell}, {Metcalfe}, {Ott}, \&
  {Schmidt}}]{pilbratt2010}
{Pilbratt}, G.~L., {Riedinger}, J.~R., {Passvogel}, T., {et~al.} 2010, \aap,
  518, L1

\bibitem[{{Rachford} {et~al.}(2002){Rachford}, {Snow}, {Tumlinson}, {Shull},
  {Blair}, {Ferlet}, {Friedman}, {Gry}, {Jenkins}, {Morton}, {Savage},
  {Sonnentrucker}, {Vidal-Madjar}, {Welty}, \& {York}}]{rachford2002}
{Rachford}, B.~L., {Snow}, T.~P., {Tumlinson}, J., {et~al.} 2002, \apj, 577,
  221

\bibitem[{{Rachford} {et~al.}(2009){Rachford}, {Snow}, {Destree}, {Ross},
  {Ferlet}, {Friedman}, {Gry}, {Jenkins}, {Morton}, {Savage}, {Shull},
  {Sonnentrucker}, {Tumlinson}, {Vidal-Madjar}, {Welty}, \&
  {York}}]{rachford2009}
{Rachford}, B.~L., {Snow}, T.~P., {Destree}, J.~D., {et~al.} 2009, \apjs, 180,
  125

\bibitem[{{Savage} {et~al.}(1977){Savage}, {Bohlin}, {Drake}, \&
  {Budich}}]{savage1977}
{Savage}, B.~D., {Bohlin}, R.~C., {Drake}, J.~F., \& {Budich}, W. 1977, \apj,
  216, 291

\bibitem[{{Seifahrt} {et~al.}(2010){Seifahrt}, {K{\"a}ufl}, {Z{\"a}ngl},
  {Bean}, {Richter}, \& {Siebenmorgen}}]{seifahrt2010}
{Seifahrt}, A., {K{\"a}ufl}, H.~U., {Z{\"a}ngl}, G., {et~al.} 2010, \aap, 524,
  A11

\bibitem[{{Sheffer} {et~al.}(2008){Sheffer}, {Rogers}, {Federman}, {Abel},
  {Gredel}, {Lambert}, \& {Shaw}}]{sheffer2008}
{Sheffer}, Y., {Rogers}, M., {Federman}, S.~R., {et~al.} 2008, \apj, 687, 1075

\bibitem[{{Snow} {et~al.}(2007){Snow}, {Destree}, \& {Jensen}}]{snow2007}
{Snow}, T.~P., {Destree}, J.~D., \& {Jensen}, A.~G. 2007, \apj, 655, 285

\bibitem[{{Sonnentrucker} {et~al.}(2007){Sonnentrucker}, {Welty}, {Thorburn},
  \& {York}}]{sonnentrucker2007}
{Sonnentrucker}, P., {Welty}, D.~E., {Thorburn}, J.~A., \& {York}, D.~G. 2007,
  \apjs, 168, 58

\bibitem[{{Sonnentrucker} {et~al.}(2010){Sonnentrucker}, {Neufeld}, {Phillips},
  {Gerin}, {Lis}, {de Luca}, {Goicoechea}, {Black}, {Bell}, {Boulanger},
  {Cernicharo}, {Coutens}, {Dartois}, {Ka{\'z}mierczak}, {Encrenaz},
  {Falgarone}, {Geballe}, {Giesen}, {Godard}, {Goldsmith}, {Gry}, {Gupta},
  {Hennebelle}, {Herbst}, {Hily-Blant}, {Joblin}, {Ko{\l}os}, {Kre{\l}owski},
  {Mart{\'{\i}}n-Pintado}, {Menten}, {Monje}, {Mookerjea}, {Pearson},
  {Perault}, {Persson}, {Plume}, {Salez}, {Schlemmer}, {Schmidt}, {Stutzki},
  {Teyssier}, {Vastel}, {Yu}, {Caux}, {G{\"u}sten}, {Hatch}, {Klein}, {Mehdi},
  {Morris}, \& {Ward}}]{sonnentrucker2010}
{Sonnentrucker}, P., {Neufeld}, D.~A., {Phillips}, T.~G., {et~al.} 2010, \aap,
  521, L12

\bibitem[{{Uttenthaler} {et~al.}(2008){Uttenthaler}, {Aringer}, {Lebzelter},
  {K{\"a}ufl}, {Siebenmorgen}, \& {Smette}}]{uttenthaler2008}
{Uttenthaler}, S., {Aringer}, B., {Lebzelter}, T., {et~al.} 2008, \apj, 682,
  509

\bibitem[{Webb \& Rao(1968)}]{webb1968}
Webb, D.~U., \& Rao, K.~N. 1968, Journal of Molecular Spectroscopy, 28, 121

\bibitem[{{Zemke} {et~al.}(1991){Zemke}, {Stwalley}, {Langhoff}, {Valderrama},
  \& {Berry}}]{zemke1991}
{Zemke}, W.~T., {Stwalley}, W.~C., {Langhoff}, S.~R., {Valderrama}, G.~L., \&
  {Berry}, M.~J. 1991, \jcp, 95, 7846

\bibitem[{{Zhu} {et~al.}(2002){Zhu}, {Krems}, {Dalgarno}, \&
  {Balakrishnan}}]{zhu2002}
{Zhu}, C., {Krems}, R., {Dalgarno}, A., \& {Balakrishnan}, N. 2002, \apj, 577,
  795

\end{thebibliography}

%%%%%%%%%%%%%%%%%%%%%%%%%%%%%%%%%%%%FIGURES%%%%%%%%%%%%%%%%%%%%%%%%%%%%%%%%%%%%%%%%%%%%%%

\clearpage
%%%%%%%%%%%%%%%%%%%%%%%%%%%%%%%%%%%%%%%%%%%%%%%%%%%%%%%%%%%%%%%%%%%%%%%%%%%%%%%%%%%%%%%%

\clearpage

%%%%%%%%%%%%%%%%%%%%%%%%%%%%%%%%%%%%%%%%%%%%%%%%%%%%%%%%%%%%%%%%%%%%%%%%%%%%%%%%%%%%%%%%

%%%%%%%%%%%%%%%%%%%%%%%%%%%%%%%%%%%%%%%%%%%%%%%%%%%%%%%%%%%%%%%%%%%%%%%%%%%%%%%%%%%%%%%%

\clearpage

%%%%%%%%%%%%%%%%%%%%%%%%%%%%%%%%%%%%TABLES%%%%%%%%%%%%%%%%%%%%%%%%%%%%%%%%%%%%%%%%%%%%%%%

%%%%%%%%%%%%%%%%%%%%%%%%%%%%%%%%%%%%%%%%%%%%%%%%%%%%%%%%%%%%%%%%%%%%%%%%%
\clearpage
\begin{longtable}{llc}
\caption{Log of Observations}\\
            &        & Integration Time \\
Target  & Date & (s)     \\
\hline
\hline
\endhead
\hline
\multicolumn{3}{p{2.5in}}{{\bf Notes:} The telluric standard star HR~6508 was observed on 2012 Aug 02 with an integration time of 336~s.}
\endlastfoot
HD 154368     & 2012 Jun 26  & 1200  \\
HD 154368     & 2012 Jul 04   & 1200 \\
HD 154368     & 2012 Aug 02 & 1440  \\
$\chi$ Oph     & 2012 Jul 04   & 240 \\
HD 149404     & 2012 Jul 04   & 600 \\
$\zeta$ Oph   & 2012 Jul 04   &  176  \\
HD 152236     & 2012 Jul 04   & 240 \\
Elias 29           & 2012 Jul 04   & 3600  \\
AFGL 2136 IRS 1      & 2012 Jul 06   & 3960 \\
HD 179406     & 2012 Jul 12   & 1200
\label{tbl_obs}
\end{longtable}

%%%%%%%%%%%%%%%%%%%%%%%%%%%%%%%%%%%%%%%%%%%%%%%%%%%%%%%%%%%%%%%%%%%%%%%%%
%\clearpage
\small
\begin{longtable}{clccccccc}
\caption{Absorption Line Parameters}\\
\hline
\hline
 &        &         & $v_{\rm LSR}$  & FWHM              & $W_{\lambda}$          & $\sigma(W_{\lambda})$ & $N(J)$                         &  $\sigma(N)$ \\
Target & Molecule & Transition & (km~s$^{-1}$) & (km~s$^{-1}$) & ($10^{-7}$~$\mu$m) & ($10^{-7}$~$\mu$m)   & ($10^{13}$~cm$^{-2}$) & ($10^{13}$~cm$^{-2}$) \\
\hline
\endhead
\hline
\multicolumn{9}{p{6.0in}}{{\bf Notes:} Columns 4 and 5 are the line-center velocity and velocity full-width at half-maximum (including instrumental broadening effects) found by a Gaussian fit to the absorption feature.  For non-detections of the $R(0)$ line a FWHM of 3~km~s$^{-1}$ is adopted in determining the uncertainty in the equivalent width.  For non-detections of the $R(1)$ line, the observed FWHM of the $R(0)$ line is adopted in determining the uncertainty in the equivalent width.  Columns 6 and 7 are the equivalent width, $W_\lambda$, and its $1\sigma$ uncertainty.  Columns 8 and 9 are the column density in the lower state, $N$, and its $1\sigma$ uncertainty.  HF column densities were derived in the optically thin limit using transition dipole moments of $|\mu_{R(0)}|^2=0.0104$~D$^2$ and $|\mu_{R(1)}|^2=0.00693$~D$^2$ \citep{zemke1991}, and transition wavelengths of 2.499385~$\mu$m and 2.475876~$\mu$m \citep{webb1968} for the $R(0)$ and $R(1)$ transitions, respectively.  Note the difference in units for the HF and H$_2$O column densities.
}
\endlastfoot
$\chi$ Oph         & HF & $R(0)$ & ...     & 3    & ...     & 0.34 & ...     & 0.31 \\
HD 149404         & HF & $R(0)$ & ...     & 3    & ...     & 0.24 & ...     & 0.22 \\
HD 152236         & HF & $R(0)$ & ...     & 3    & ...     & 0.37 & ...     & 0.34\\
HD 154368         & HF & $R(0)$ & 4.9  & 3.7 & 1.80 & 0.41 & 1.66 & 0.37 \\
HD 154368         & HF & $R(1)$ & ...  & ...        & ...    & 0.17 & ...     & 0.23 \\
Elias 29               & HF & $R(0)$ & 3.0  & 4.0 & 19.5 & 0.70 & 18.0 & 0.65 \\
Elias 29               & HF & $R(1)$ & ...  & ...        & ...    & 0.19 & ...     & 0.27 \\
AFGL 2136 IRS 1 & HF & $R(0)$ & 20.3 & 11.6 & 26.2 & 1.64 & 24.2 & 1.52 \\
AFGL 2136 IRS 1 & HF & $R(1)$ & 24.5 & 14.8 & 17.8 & 3.10 & 24.9 & 4.34 \\
\hline
 &        &         & $v_{\rm LSR}$  & FWHM              & $W_{\lambda}$          & $\sigma(W_{\lambda})$ & $N(J_{K_{a}K_{c}})$                         &  $\sigma(N)$ \\
Target & Molecule & Transition & (km~s$^{-1}$) & (km~s$^{-1}$) & ($10^{-7}$~$\mu$m) & ($10^{-7}$~$\mu$m)   & ($10^{16}$~cm$^{-2}$) & ($10^{16}$~cm$^{-2}$) \\
\hline
AFGL 2136 IRS 1 & H$_2$O & $\nu_1$~9$_{6,3}$--8$_{5,4}$   & 23.7 & 13.4 & 30.0 & 2.04 & 3.38 & 0.23 \\
AFGL 2136 IRS 1 & H$_2$O & $\nu_1$~10$_{7,4}$--9$_{6,3}$ & 24.8 & 17.6 & 21.1 & 3.36 & 2.43 & 0.39
%HD 179406         & $R(0)$ & ...   &  &  &  &  &  & 1 
\label{tbl_absorption}
\end{longtable}
\normalsize

%%%%%%%%%%%%%%%%%%%%%%%%%%%%%%%%%%%%%%%%%%%%%%%%%%%%%%%%%%%%%%%%%%%%%%%%%
%\clearpage
\begin{longtable}{lccccccc}
\caption{HF and H$_2$ Abundances}\\
            &  $N({\rm HF})$               & $N({\rm H}_2)$               & H$_2$       & $N({\rm CH})$                & CH
            & $N({\rm HF})/N({\rm H}_2)$ &  $N({\rm HF})/N({\rm CH})$ \\
Target  & ($10^{13}$~cm$^{-2}$) &  ($10^{21}$~cm$^{-2}$) & Reference & ($10^{13}$~cm$^{-2}$) & Reference
            & ($10^{-8}$)                          &  \\
\hline
\hline
\endhead
\hline
\multicolumn{8}{p{6.0in}}{{\bf Notes:} Upper limits on $N({\rm HF})$ in a single cloud component are taken to be 3 times the 1$\sigma$ uncertainties reported in Table \ref{tbl_absorption}.  Upper limits on the total line-of-sight column densities of HF (column 2) account for multiple clouds (velocity components) observed in CH \citep[3 components for HD~149404, and 2 for HD~152236;][]{gredel1993}, and are used in calculating upper limits to the HF/H$_2$ and HF/CH abundance ratios.   Upper limits on the $J=1$ column densities toward HD~154368 and Elias~29 are small compared to observed $J=0$ column densities so we set the total HF column densities equal to the latter.  For AFGL~2136~IRS~1 we present results for multiple cases described in this paper.  The observed line-of-sight column densities (row labeled LoS) are presented, where the HF column density is taken to be the sum of the columns in the $J=0$ and $J=1$ levels and the H$_2$ column density is that reported by \citet{kulesa2002}.  Rows labeled ``warm'' and ``cold'' present the results from our analysis in Section \ref{subsec_2136} where we assume both warm and cold components are present.  This decomposition into warm and cold gas components is likely the more realistic picture.\newline
{\bf References:} (1) \citet{savage1977}; (2) \citet{rachford2009}; (3) \citet{rachford2002} ; (4) \citet{kulesa2002} (Note that only the {\it para}-H$_2$ column density was measured---via the $v=1$--0, S(0) transition---and that the total H$_2$ column density was estimated assuming thermal equilibrium between {\it ortho} and {\it para} H$_2$, and a kinetic temperature derived from CO observations also presented therein); (5) Section \ref{subsec_2136} herein; (6) D.~E.~Welty (2002, private communication)
}
\endlastfoot

$\chi$ Oph         & $<0.93$            & $0.43\pm0.10$ & 1 & $3.02\pm0.21$ & 6 &  $<2.18$ & $<0.31$ \\
HD 149404         & $<1.99$            & $0.62\pm0.05$ & 2 & $2.88\pm0.20$ & 6 &  $<3.23$ & $<0.69$ \\
HD 152236         & $<2.02$            & $0.53\pm0.13$ & 2 & $2.40\pm0.20$ & 6 &   $<3.79$ & $<0.84$ \\
HD 154368         & $1.66\pm0.37$ & $1.44\pm0.40$ & 3 & $6.31\pm0.44$ & 6 &  $1.15\pm0.41$ & $0.26\pm0.06$ \\
Elias 29               & $18.0\pm0.65$ & $26\pm5$         & 4 & ... & ...                 & $0.69\pm0.14$ & ... \\
AFGL 2136 (LoS) & $49.1\pm4.60$ & $56\pm9$         & 4 & ... & ...                 &  $0.88\pm0.16$ & ... \\
AFGL 2136 (warm) & $162\pm28.2$  & 55--98              & 5 & ... & ...                 &  1.65--2.94 & ... \\
AFGL 2136 (cold) & $15.0\pm2.21$ & 26--46              & 5 & ... & ...                 &  0.33--0.58 & ... 
\label{tbl_ratios}
\end{longtable}

\end{document}